# Analysis of magnetization induced error field in superconducting magnets

Jun Lu

*National High Magnetic Field Laboratory, Florida State University, Tallahassee Florida 32310, USA*

**ABSTRACT**

The field error due to magnetization of superconductors in a superconducting magnet can significantly compromise the field homogeneity which is extremely important for NMR magnets. This is especially true for high $T_c$ superconducting magnets. In this paper, the error field induced by the magnetization of superconducting wires or tapes is analyzed using a magnetic dipole field model. It uses experimentally measured magnetization of superconducting wires, and eludes the need of assuming screening current and its distribution as in previous methods. The field error of a REBCO coil calculated by this model and by a screening current model shows complete agreement. A few cases of field error distribution along the coil axis are calculated. The results are in qualitative agreement with experiments in the literature. This model would be particularly useful for magnets wound by multifilamentary wires or tapes where the filament sizes vary and filament bridging may occur hence screening current distribution is unknown. This model can also be used as a base for numerical calculations of the field error of a large practical magnet.

1. Introduction

Magnetic field produced by flowing an electrical current through a superconducting coil usually shows a small hysteresis. This hysteresis is due to the magnetization of the superconducting wires originated from screening currents in them. The magnitude of this additional field, or the error field, depends on the properties of the superconducting wire as well as the size of the coil. The remnant fields due to the remnant magnetization was reported in the early days of superconducting magnet development[1,2]. M.A. Green analyzed the problem by using a so-called screening current 'doublet' model[3] which was later used for calculating the error field of Tevatron magnets[4]. A method for mitigating the error field was suggested and demonstrated by Misek and Svoboda[5], which has been adopted in commercial magnet systems[6].

The error field is relatively small in magnets made with multifilamentary wires with fine filaments such as NbTi, For magnets wound with high critical current density Nb$_3$Sn wires, the error field from wire magnetization is significantly larger[7,8]. The error field is even larger in high temperature superconductor (HTS) magnets especially REBCO magnets[9-16]. The error field as well as its reduction becomes an important topic for high field HTS magnet development community[17-20]. So far most error field analyses use a screening current induced field (SCIF) model in REBCO magnets. They are numerical analyses providing detailed information about SCIF for specific REBCO coils. However, many aspects of the SCIF is still to be understood partly due to the uncertainty in the magnitude and distribution of the screening currents in REBCO. In addition, a reliable method for calculating the error field in magnets made by mulitifilamentary superconducting wires or tapes is needed.

I will analyze the error field by using a magnetic dipole field model. For clarity, below I will use the term magnetization-induced-error-field (MIEF) to describe the error field instead of SCIF. This is to emphasize the view point of the magnetization instead of screening current. The MIEF is analyzed by using a simple magnetic dipole model which can be directly linked to the experimentally measured magnetization of superconductors. The model provides a simpler physical picture of this phenomenon, and could serve either in a preliminary magnet design calculation or as a base for more elaborate full-scale design calculations.

## 2. The model

*2.1 Magnetization of superconductors*

The origin of the MIEF is the magnetization of superconducting wires. The magnetization of a few practical superconducting wires is plotted as a function of applied magnetic field in Fig. 1. For better presentation, the REBCO magnetization is reduced by a factor of 5. The data were measured by a vibrating sample magnetometer in a physical property measurement system made by Quantum Design Inc. USA. Applied field is perpendicular to NbTi and $Nb_3Sn$ round wire longitudinal direction, and the broad face of Bi-2223 and REBCO tapes. Table I lists some relevant parameters of these samples.

TABLE I. Superconductor wires tested by magnetometry

|  | NbTi | $Nb_3Sn$ | Bi-2223 | REBCO |
| --- | --- | --- | --- | --- |
| ID | ITER PF | B-OST Hi-Lumi | Sumitomo HT-NX | SuperPower |
| Wire size (mm) | ϕ 0.73 x 100 | ϕ 0.85 x 100 | 0.268 x 4.45 x 2.4 | 0.15 x 4.1 x 6.7 |
| Sample shape | 7 turn coil | 7 turn coil | Short piece | Short piece |

For a short REBCO sample, the measured magnetization $M$ depends on its length following[21]

$$M = (1/4)J_c[w(1 - w/3l)] \quad (1)$$

Where $J_c$ is the critical current density, $l$ the sample length, $w$ the sample width. Let the magnetization of a long sample $M_0 \equiv (1/4)J_c w$, then

$$M = M_0 (1-w/3l) \quad (2)$$

The magnetization of REBCO in Fig. 1 is $M_0$ corrected for the short sample length by (2) and divided by 5 for easy comparison with those of other superconductors. Evidently when the applied field is perpendicular to its broad face, REBCO has much larger magnetization than other superconductors.

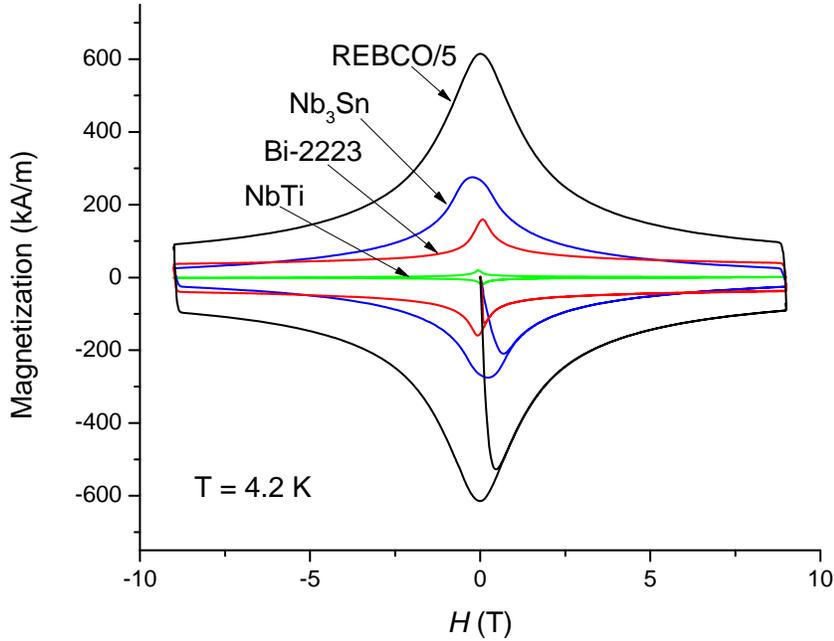

FIG. 1. The magnetization of a few practical superconducting wires as a function of magnetic field at 4.2 K with field perpendicular to the wire longitudinal direction. For Bi-2223 and REBCO tape samples, the field was perpendicular to the sample's broad face. REBCO data is reduced by a factor of 5 for better presentation.

*2.2 Magnetic dipole moment model*

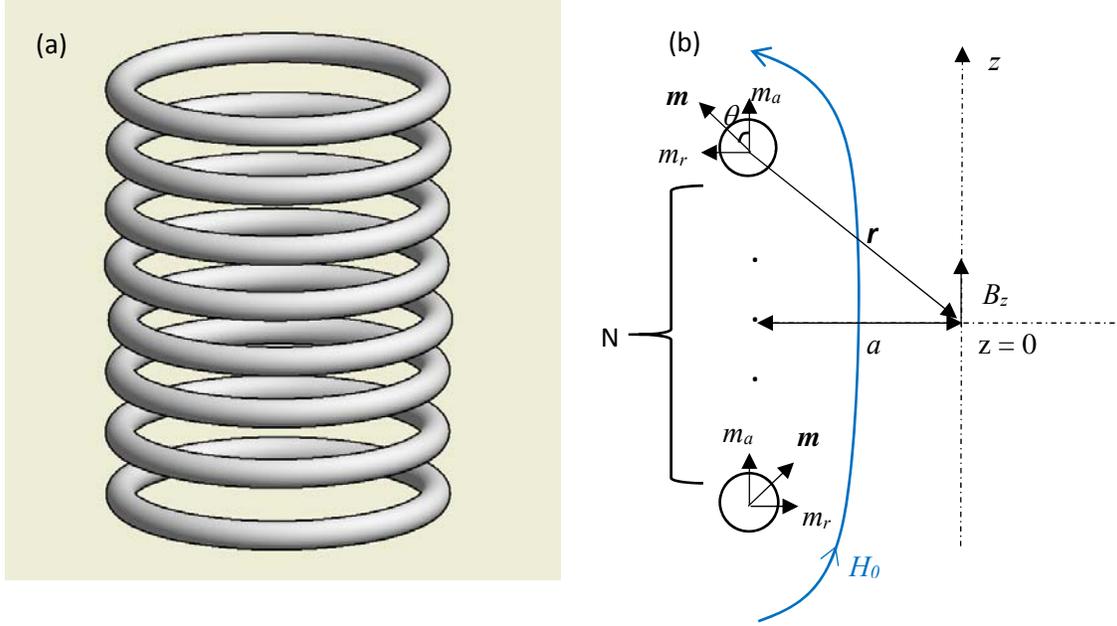

FIG.2. (a) A schematic drawing of a single-layer superconducting coil. (b) Coordinates used in the model: $m$ is the magnetic moment vector, z the axial direction with coil mid-plane at $z = 0$, $r$ the displacement vector from the wire to a point of interest on z axis, $\theta$ the angle between $r$ and z directions, $a$ the coil radius.

A magnetic dipole field model is used in the following analyses. Fig. 2 (a) is a sketch of a superconducting coil. The coordinates and symbols used below is depicted in Fig. 2(b), where the magnetic field vector $H_0$ magnetizes the coil, and the induced magnetic moment $m$ at a displacement $r$ contributes to the error field $B_z$ along the coil axis z. The coil radius is $a$. The coil mid-plane is at z = 0. When $r$ is much greater than the width of the wire, the field $dB$ generated by a magnetic dipole $dm$ is[22]

$$d\boldsymbol{B}(\boldsymbol{r}) = \frac{\mu_0}{4\pi}\frac{3\hat{r}(\hat{r}\cdot d\boldsymbol{m})-d\boldsymbol{m}}{|r|^3} \qquad (3)$$

Where $\hat{r}$ is the unit vector of $r$. When $dB$ is integrated over a coil, the radial component of the integrated field is zero due to the symmetry. Therefore only the axial dipole field $dB_z$ needs to be considered. $m$ can be disassociated to a radial and an axial component $m_r$ and $m_a$, whose respective contributions to $dB_z$ are $dB_{zr}$ and $dB_{za}$. So we have

$$dB_{zr}(z) = \frac{\mu_0}{4\pi}\frac{3\sin\theta|dm_r|\cos\vartheta}{|r(z)|^3} = \frac{\mu_0}{4\pi}\frac{3az|M|}{|a^2+z^2|^{5/2}}dv \qquad (4)$$

$$dB_{za}(z) = \frac{\mu_0}{4\pi} \frac{|dm_a|[-3\cos^2\vartheta - 1]}{|r(z)|^3} = \frac{\mu_0}{4\pi} \frac{|M|(-(3z^2)/(a^2+z^2)-1]}{|a^2+z^2|^{3/2}} dv \qquad (5)$$

Where $\theta$ is as depicted in Fig. 1. $|M|$ is the magnitude of magnetization, $dv$ is the volume of the superconductor segment. The integrated magnetic dipole field on the coil axis, i.e. the on-axis MIEF, is the integral $dB_z$ over the entire coil,

$$B_z = \int dB_{zr} + \int dB_{za} \qquad (6)$$

Once the magnetization of the superconductor is measured separately, the MIEF of superconducting coils can be calculated by equations (4)-(6). For off-axis MIEF of solenoids or other types of magnets, such as dipole and quadrupole magnets, equation (2) is still valid, but the calculation will be more complicated.

## 3   Results and discussions

*3.1 Loops of round wires*

Let us start with a simplest case of a single loop of round wire. It is assumed that the orientation of the magnetization is at an angle $\theta$ with respect to the axial direction z of the loop. For a loop of $a$ = 20 mm wound with 1 mm diameter Nb$_3$Sn wire which has magnetization of 270 kA/m, the MIEF $B_z$ are calculated with angles $\theta$ = 0°, 45° and 90°, the results are plotted in Fig. 3. It shows that when the magnetization is parallel to z ($\theta$ = 0°), $B_z$ is symmetrical about the coil plane at z = 0. When the dipole moment is perpendicular to z ($\theta$ = 90°), $B_z$ has a mirror symmetry and change the polarity at z = 0. The case of $\theta$ = 45° or other arbitrary angles is just the weighted sum of the two. This means that the MIEF of a one loop could be asymmetrical about the loop plane z = 0, which seems to be inconsistent with what were observed in multiple-turn magnet coils.

So next, let us look at the case of two loops. Each loop is the same as the single loop above but separated by 60 mm in z direction, one at z = 30 mm and the other at z = -30 mm. It is also assumed that when the magnetization is at an angle $\theta$ for the top loop, it is at $-\theta$ for the bottom loop, as is typical for the top and bottom turns of a multi-turn solenoid. The MIEF in this case is symmetric about z = 0 as shown in Fig. 4. The magnitude and axial distribution of the MIEF strongly depends on $\theta$, especially near z = 0. For a practical multi-turn coil, the MIEF is obtained by integrating over all the turns. It is conceivable that the general feature of $B_z(z)$ in a practical coil could be similar to those in Fig. 4, as it is noted that the curves in Fig. 4 shows similar characteristics as the experimental data of remnant field in superconducting magnets in the literature[1,5,6,10,12,19,20].

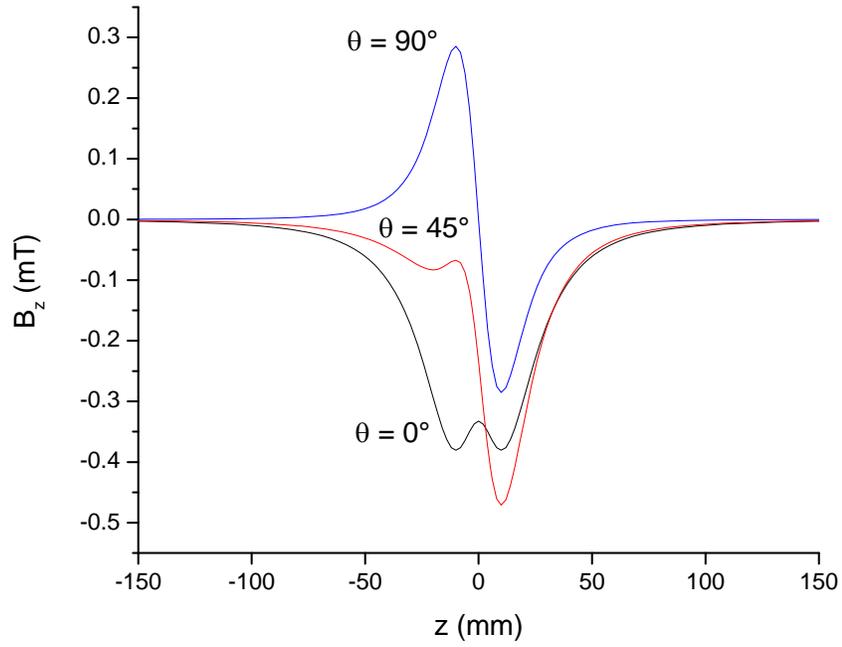

FIG. 3. The MIEF of a single loop as a function of z. The coil radius *a* is 20 mm, the wire is 1 mm diameter with magnetization M = 270 kA/m. Three curves are for magnetization is parallel ($\theta = 0°$) and perpendicular ($\theta = 90°$), and $\theta = 45°$ to the loop axial direction respectively.

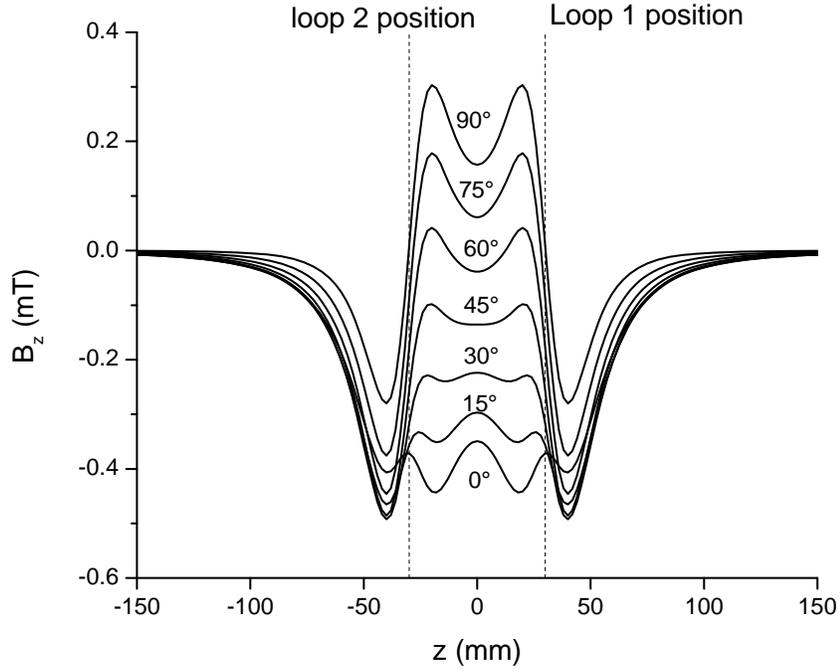

FIG. 4. MIEF of two superconducting loops located at z = 30 mm and -30 mm respectively. Each is the same as the one in Fig. 3. The orientation of the magnetization with respect to the axial direction z is indicated on each curve. It is assumed that when one loop is at $\theta$, the other is at $-\theta$.

*3.2 Loops of REBCO tape*

MIEF in REBCO coils is the most important case from application point of view. This is because, the magnetization in REBCO is much greater, so the MIEF in REBCO coils is much larger and has to be dealt with for most of applications. Another difference between the round wire and the REBCO tape is that the magnetization parallel to REBCO broad face is negligibly small, as shown experimentally in ref. 23. Therefore in the MIEF calculation of REBCO coils only the perpendicular magnetization is considered.

I calculated a case of 12 turn single layer REBCO coil with radius of 20 mm and no gap between turns. The REBCO tape is 4 mm wide and 0.15 mm thick. The magnetization is 3100 kA/m. The magnetization moment for the turns above the mid-plane has $\theta = 90°$, below the mid-plane has $\theta = -90°$. If the perpendicular (radial) magnetization can only be induced by the radial field, then it will be zero near the coil's mid-plane where the radial field is zero. This implies that most MIEF is from the magnetization at the end turns. In order to cover various radial field distributions, MIEF is calculated for three cases. A) all the turns have full strength of magnetization, B) only two end turns have full magnetization, the rest of the turns have zero magnetization, and C) the magnetization has a parabolic distribution long z axis, from 0 at the mid-plane to 100% at the end turns. The calculated results are shown in Fig. 5 where the mid-plane is at z = 0, and the ends of the coil is at -24 mm and 24 mm respectively. It is noted that the MIEF is positive near the mid-plane and negative at the coil ends.

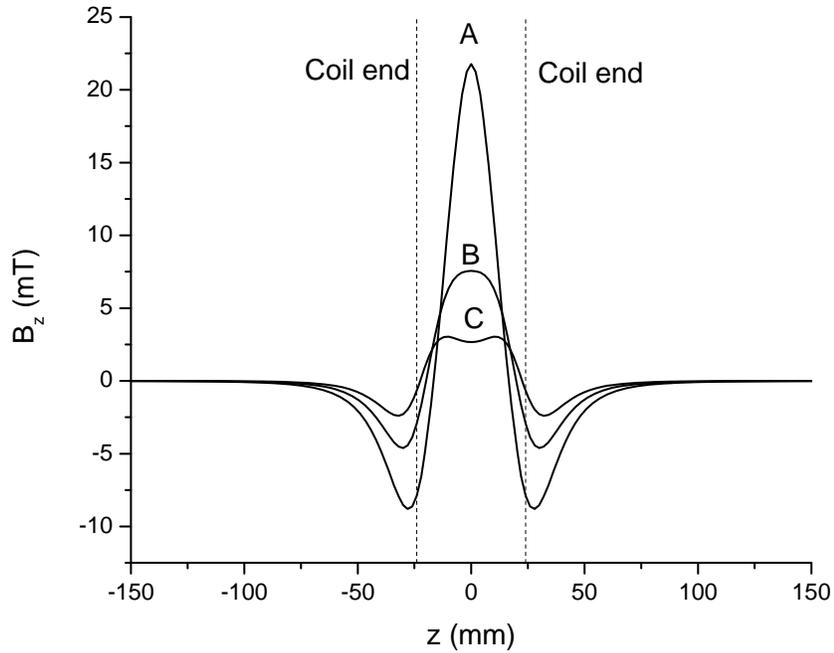

FIG. 5. The MIEF axial profile of a 12 turn single layer REBCO coil for three cases. A, full magnetization of all the turns. B, full magnetization only at two end turns. C, magnetization increases from the mid-plane to end turns following a parabolic function.

*3.3 Discussions*

Equation (1) is a far-field approximation. When the distance is comparable to wire diameter or width, inaccuracy may occur. So in order to verify the accuracy of above calculations, a screening current model is used to calculate the SCIF in the 12 turn REBCO coil and to compare with the calculated MIEF. In the SCIF model, for each turn two screening current loops flowing in opposite directions and separated by 2 mm is assumed. The screening current is assumed to be 930 A, which is obtained from magnetization of 3100 kA/m and equation (1). For a coil with 5 mm radius, which would be the smallest inner radius of a practical magnet, the calculated results by the two models show complete agreement with each other. This verifies that in REBCO coils the MIEF model is equivalent to the SCIF model used by other researchers.

During the operation of a superconducting magnet, the transport current coexists with the screening currents. It was derived[24] based on Bean's critical state model that in magnetic fields above the penetration field, magnetization is reduce by a factor of $1 - (I_t/I_c)^2$ due to transport current $I_t$, where $I_c$ is the critical current. Although a direct experimental proof of this magnetization reduction seems to be difficult to obtain, it appears reasonable that when the transport current occupies a significant fraction of the wire cross-section, the screening currents which share the cross-section with the transport current have to be reduced. As a consequence, the magnetization decreases with increasing transport current.

The MIEF is detrimental to the magnet field homogeneity, therefore it is highly desirable to reduce it especially for REBCO magnets. Because the magnetization is proportional to the width of tapes or wire filaments, reducing the filament size or tape width should proportionally reduce the MIEF. Another method commonly used to reduce the magnetization effect is to demagnetize the superconductors by applying an oscillatory field with a damping amplitude, analogous to the method for demagnetizing a ferromagnetic object. This method has been used for minimize the MIEF in NbTi or $Nb_3Sn$ magnets such as those in Quantum Design's PPMS, and has been experimented in REBCO magnets with promising results[17-19]. Another related issue that is harmful to REBCO magnet field quality is the temporal drift in the power driven mode. This drift is a direct consequence of drift in the MIEF due to a slow decay in the magnetization. The drift in the magnetization can be attributed to the phenomenon of flux creep, and it can be measured experimentally on a small sample by magnetometry.

For a large coil, the MIEF includes contributions from all the turns. So its magnitude and polarity depends on the distribution of the magnetization, which in turn depends on the magnitude and the history of the primary field which magnetize the conductors. According to equation (3), MIEF decreases quickly with increasing distance ($\sim r^{-3}$). Therefore for a large coil, only those magnetized turns near the magnet center make significant contribution to the MIEF at the center where the field quality is of interest. So it can be speculated that the MIEF increases with coil size slowly. It is likely that the main field of a solenoid increases with its size faster than the increase of its MIEF, so the relative effect of the MIEF (MIEF to main field ratio) is smaller as the coil becomes larger and its central field becomes higher. This is consistent with the reports that an appreciable MIEF was found in a small REBCO coil[14], while a relatively small MIEF was observed in a large REBCO coil[25]. In addition, if significant magnetization only occurs at end turns of a REBCO coil where the primary radial field is significant, the taller the coil the smaller the MIEF.

The MIEF model uses measured magnetization of the wires/tapes without making assumptions on screening currents. Compared with the SCIF model, the MIEF model provides an easier approach to the field error calculation especially in a magnet made of mutifilamentary wires or tapes where screening currents and their distributions are not well-defined and difficult to verify experimentally. Moreover with MIEF model, phenomena such as MIEF reduction by field oscillation and the MIEF drift due to flux creep can be understood and dealt with more easily.

## 4  Conclusion

The MIEF in superconducting coils is calculated using magnetic dipole field from superconducting wires. The calculation is based on experimentally measureable magnetization of superconducting wires without the assumptions for screening currents and its distributions. The MIEF model is used to calculate the error field of a REBCO coil and compared it with that by a SCIF model. The two agree with each other very well. A few cases of MIEF distribution along the coil axis are calculated and the results are qualitatively agree with experiments in the literature. Since the assumption of screening currents is not needed, this model would be particularly useful for cases where screening current distribution is not well-defined, such as in multifilamentary wires/tapes where the filament sizes vary and filament bridging may occur. Furthermore, this model can be used as a base for numerical calculations of error fields of large practical magnets.

## 5  Acknowledgement

This research is carried out at the NHMFL which is supported by NSF through NSF-DMR-1157490 and 1644779, and the State of Florida.